\documentclass[aps,pre,twocolumn,superscriptaddress,secnumarabic,showpacs,floatfix,amsmath,amssymb]{revtex4}
\usepackage{latexsym}
\usepackage{amsmath}
\usepackage{amssymb}
\usepackage{amsfonts}
\usepackage{graphicx}

\begin{document}

\title{Subdiffusive behavior in  a trapping potential:
mean square displacement and velocity autocorrelation function}

\author{M. A. Desp\'osito}

\email[]{mad@df.uba.ar}

\affiliation{Departamento de F\'{\i}sica, Facultad de Ciencias
Exactas y Naturales,\\ Universidad de Buenos Aires, 1428 Buenos
Aires, Argentina.}

\affiliation{Consejo Nacional de Investigaciones Cient\'{\i}ficas
y T\'{e}cnicas, Argentina.}

\author{A. D. Vi\~{n}ales}

\affiliation{Departamento de F\'{\i}sica, Facultad de Ciencias
Exactas y Naturales,\\ Universidad de Buenos Aires, 1428 Buenos
Aires, Argentina.}


\date{\today}


\begin{abstract}

A  theoretical framework for analyzing stochastic data from single-particle tracking
in complex or viscoelastic
materials and  under the influence of a trapping potential is presented.
Starting from a generalized Langevin equation we found explicit expressions for the
two-time dynamics of the tracer particle. The mean square displacement and the velocity
autocorrelation function of the diffusing particle
are given in terms of the time lag.
In particular, we investigate the subdiffusive case. The exact solutions are discussed
and the validity of usual approximations are examined.

\end{abstract}




\maketitle


\section{Introduction}

The viscoelastic properties of complex fluids, like  polymers, colloids and biological materials,
can be derived from the dynamics of individual spherical particles embedded in it \cite{Wai,Val}.
Particle tracking microrheology experiments \cite{Wai,Val,Yam,Lau,Ma95,Ma97,Qian,Lev} are based on the
observation of the motion of individual tracer particles.
In a typical microrheology experiment, particle positions are recorded in the form of a
time sequence  and information about the dynamics is essentially extracted  by measuring
the mean square displacement of the probe  particles  \cite{LuSol,Dan,Gold}.
Based on a generalized Langevin equation with a memory function and assuming that inertial
effects are negligible, Mason and coworkers \cite{Ma95,Ma97} have obtained a direct relation
between the mean-square displacement of  free
tracer particles and the viscoelastic parameters of the environment.
It has been recently noted that the fluid inertia and the resulting
memory effects become increasingly important
when high-resolution experiments are performed \cite{Luk1}.


On the other hand, optical traps are increasingly used
for position detection, with a wide range of
applications in physics and biology \cite{Yam,Neu,Luk2,Ata,Fis,Cas}.
In an optical trap, the interaction between the laser and the
trapped object can be approximated by a harmonic potential \cite{Den}.
However, the use of a trapping  complicates the analysis of
the  obtained data, since  the interactions with the
viscoelastic environment overlaps with the influence of the
trapping force \cite{Wai,Luk1}.
For example, it has been  noticed  that neglecting memory effects leads to
calibration errors of optical traps \cite{Luk2}.


It is now well established that when particles diffuse through a soft complex fluids
or  biological materials,  they exhibit anomalous subdiffusive  behaviors \cite{Val,Dan,Bra,Gui,Tol}.
In this situation, the mean-square displacement exhibit  a slow relaxation
with the presence power-law decay in the
range of large times.
A theoretical complete description of the behavior of a particle in a complex medium
and subjected to a harmonic potential can be formulated in terms of the
generalized Langevin equation (GLE) containing a memory function \cite{WaTo,Porr,Pot,VD}.
In a recent paper \cite{VD}, we have obtained  analytical expressions for the evolution
of mean values and variances in terms of Mittag-Leffler functions.
However, from an experimental point of view it is necessary to get expressions
for the two-times correlation functions. For instance,
the mean square displacement (MSD) can be expressed  as
\begin{eqnarray}
\rho(\tau)=  \lim_{ t \to \infty } \langle \left(X(t+\tau)-X(t)\right)^{2} \rangle
\label{defmsd} \, ,
\end{eqnarray}
where $|X(t+\tau)-X(t)|$ is the particle displacement between two time points,
 $t$  denote  the \textit{absolute time} while  $\tau$ is the so-called
\textit{lag time} \cite{Metz}.
Alternative information about the experimentally observed diffusive behavior can be
extracted from the normalized velocity autocorrelation function (VACF) \cite{Qian1},
defined as \cite{Porr,VD}
\begin{eqnarray}
C_V(\tau)=  \lim_{ t \to \infty } \frac{\langle V(t+\tau)V(t)\rangle}{\langle V^{2}(t)\rangle} \, .
\label{defvacf}
\end{eqnarray}
Then,  to calculate the MSD and the VACF  one must know the behavior of the two-time
correlations $\langle X(t+\tau)X(t)\rangle$ and $\langle V(t+\tau)V(t)\rangle$.
In what follows we will investigate the  behavior of the MSD (\ref{defmsd})  and VACF (\ref{defvacf})
for harmonically bounded particle immersed in a viscoelastic environment.
For this purpose, in Sec. 2 we present the  corresponding generalized Langevin equation (GLE).
The two-time dynamics is obtained, which enable us to calculate the MSD and VACF for arbitrary memory kernels.
Section 3 is devoted to the study of the  subdiffusive case. The analytical solutions
are given and  compared with the overdamped approximation.
Finally, a Summary of our results  is presented in Sec.4 .


\section{Diffusion in a harmonic well}

\subsection{Formal solution for the GLE}

In what follows we consider the dynamics of a test particle of mass $m$,
immersed in a complex or  viscoelastic environment   and simultaneously bounded
in a harmonic potential well.
The resulting motion can be  described by the following GLE
\begin{eqnarray}
m \ddot{X}(t) + m \int_0^t dt' \gamma (t-t') \dot{X}(t') +  m \omega_0^2 X  = F(t) \,
, \label{Lang}
\end{eqnarray}
where $\omega_0$ is the frequency of the trap, $\gamma(t)$ is the dissipative memory kernel,
and  the internal noise $F(t)$ is a zero-centered and stationary random force with
correlation function
\begin{eqnarray}
\langle F(t) F(t') \rangle = C(|t - t'|)=C(\tau)
 \, .\label{rint}
\end{eqnarray}

The integral term in (\ref{Lang}) represents  the dependence  of the viscous force on the velocity
history and the memory kernel  $\gamma(t)$ is related to the noise correlation function $C(t)$
via the second fluctuation-dissipation theorem \cite{Zwa}
\begin{eqnarray}
C(t)= k_B T \, m\,\gamma(t)
 \, ,\label{tfd}
\end{eqnarray}
where $T$ is the absolute temperature, and $ k_B $ is the Boltzmann
constant.


In what follows we consider the  one-dimensional case, but our  results can be easily extrapolated
to the  two or three dimensional case.
The Langevin equation (\ref{Lang}) can be  formally solved by means
of the Laplace transformation.
Taking into account the deterministic initial conditions $x_0 = X(0)$
and $v_0 = \dot{X}(0)$, the evolution of the Laplace transform of the position $X(t)$ reads
\begin{eqnarray}
\widehat{X}(s) = x_0\left(\frac{1}{s}-\omega_0^2\,\widehat{I}(s)\right)
+  \left(v_0 +  \frac{1}{m}\widehat{F}(s)
\right)\widehat{G}(s) \,  , \label{Xs}
\end{eqnarray}
where $\widehat{F}(s)$ is the Laplace
transform of the noise.
The  relaxation function  $G(t)$ is the  Laplace inversion of
\begin{eqnarray}
\widehat{ G}(s) =  \frac{1}{s^2 + s\widehat{\gamma}(s)+\omega_0^2} \,  ,
 \label{Gs}
\end{eqnarray}
where $\widehat{\gamma}(s)$ is the Laplace transform of the
damping kernel, and
\begin{eqnarray}
 \widehat{I}(s) = \frac{ \widehat{G}(s)}{s} \, ,
 \label{IGs}
\end{eqnarray}
is the Laplace transform of
\begin{eqnarray}
 I(t) =  \int_0^t dt' G(t')\, .
 \label{relaIG}
\end{eqnarray}

On the  other hand, the Laplace transform of the velocity $V(t)= \dot{X}(t)$ satisfies that
\begin{eqnarray}
\widehat{V}(s) = \left(v_0 +  \frac{1}{m}\widehat{F}(s)
\right)\widehat{g}(s) - x_0\,\omega_0^2\,\widehat{G}(s)\label{Vs} \,   ,
\end{eqnarray}
where
\begin{eqnarray}
\widehat{g}(s) =  s \, \widehat{ G}(s).
 \label{Is}
\end{eqnarray}

From Eqs. (\ref{Xs}) and (\ref{Vs}) a formal expression for the displacement $X(t)$ and the velocity $V(t)$
can be written as
\begin{eqnarray}
X(t) &=&  \langle X(t) \rangle  + \frac{1}{m}\int_0^t dt' G(t-t')
F(t') \, , \label{X} \\
V(t) &=&  \langle V(t) \rangle  + \frac{1}{m}\int_0^t dt' g(t-t')
F(t') \, , \label{V}
\end{eqnarray}
where
\begin{eqnarray}
 \langle X(t) \rangle &= &  x_0\left(1-\omega_0^2 \, I(t)\right) + v_{0} \, G(t)  \,  ,
 \label{VMX} \\
  \langle V(t) \rangle &= &  v_0 \, g(t) - x_0\,\omega_0^2\, G(t) \,  ,
 \label{VMV}
\end{eqnarray}
are the position and velocity mean values evolution, respectively.

\subsection{Expressions for the MSD and VACF}

To calculate the  two-time properties of the dynamical variables
involved in the expressions of the MSD (\ref{defmsd})  and VACF (\ref{defvacf})
we will make use of the double Laplace transform technique \cite{Pot}.
Then, from (\ref{Xs}) and  (\ref{Vs}) we have
\begin{eqnarray}
\langle\widehat{X}(s)\widehat{X}(s')\rangle & =& x_0^{2} \, \widehat{\chi}(s)\widehat{\chi}(s')   +
v_0^{2} \,\widehat{G}(s)\widehat{G}(s')   \nonumber \\
& +&   x_0 \, v_0 \,
(\widehat{\chi}(s)\widehat{G}(s')+\widehat{\chi}(s')\widehat{G}(s)  )  \nonumber \\
&+ &
\frac{1}{m^{2}}\, \widehat{G}(s)\widehat{G}(s')\langle
  \widehat{F}(s)\widehat{F}(s') \rangle \label{Xss} \,  ,
\\
\langle\widehat{V}(s)\widehat{V}(s')\rangle & =& v_0^{2} \, \widehat{g}(s)\widehat{g}(s')   +
x_0^{2} \, \omega_0^4 \, \widehat{G}(s)\widehat{G}(s')    \nonumber \\
& -&   x_0 \, v_0\, \omega_0^2 \,
(\widehat{g}(s)\widehat{G}(s')+\widehat{g}(s')\widehat{G}(s)  )  \nonumber \\
&+ &
\frac{1}{m^{2}}\, \widehat{g}(s)\widehat{g}(s')\langle
  \widehat{F}(s)\widehat{F}(s') \rangle \label{Vss}  \,   ,
\end{eqnarray}
where
\begin{eqnarray}
 \widehat{\chi}(s) = \frac{1}{s}-\omega_0^2\widehat{I}(s) \,
 \label{chis}
\end{eqnarray}
is the Laplace transform of $\chi(t) =  1-\omega_0^2 I(t)$.


\begin{widetext}

In the Appendix  we show how the last
term of Eqs. (\ref{Xss}) and (\ref{Vss}) can be calculated.
Inserting  expressions (\ref{GGFF}) and (\ref{ggFF}) into (\ref{Xss}) and (\ref{Vss})
and making a double Laplace inversion, we arrive at
\begin{eqnarray}
\langle X(t)X(t') \rangle & =&
x_0^{2} \,  \chi(t)\chi(t')   + (v_0^{2}-\frac{k_B T }{m})
G(t)G(t')
+   x_0 \, v_0\,
(\chi(t)G(t')+\chi(t')G(t) )\nonumber \\
&+&
\frac{k_B T }{m} \left(I(t)+ I(t')-I(|t-t'|)
\right)
- \frac{k_B T }{m}
\,\omega_0^2 \, I(t)I(t') \,  ,
\label{xtxt}
\\
\langle V(t)V(t') \rangle & =&
\frac{k_B T }{m}\,  g(|t-t'|)
+(v_0^{2}-\frac{k_B T }{m})
g(t)g(t')
+\omega_0^2\, (x_0^{2}\omega_0^2-\frac{k_B T }{m})
G(t)G(t')\nonumber \\
&- & x_0 v_0 \,\omega_0^2\, (g(t)G(t')+g(t')G(t) )
\label{vtvt}
\end{eqnarray}

Finally, by considering time lags $\tau >0$, from (\ref{xtxt}) and (\ref{vtvt})  we have
\begin{eqnarray}
\langle \left(X(t+\tau)-X(t)\right)^{2} \rangle  &=& \frac{2 k_B T}{m} \,
I(\tau)
 - 2 x_0 v_0 \omega_0^2 \left(G(t+\tau)-
G(t)\right)\left(I(t+\tau)-
I(t)\right)\nonumber \\
& +&  (v_0^{2}-\frac{k_B T }{m}) \left(G(t+\tau)-
G(t)\right)^{2}
 +\omega_0^2\,(x_0^{2}\omega_0^2-\frac{k_B T }{m})\left(I(t+\tau)-
I(t)\right)^{2}
  \,    ,
\label{xcorrtau} \\
\langle V(t+\tau)V(t) \rangle & =&
\frac{k_B T }{m} \, g(\tau)
+(v_0^{2}-\frac{k_B T }{m})
g(t+\tau)g(t)
+\omega_0^2\,(x_0^{2}\omega_0^2-\frac{k_B T }{m})
G(t+\tau)G(t)
\nonumber \\
&-& x_0 \,v_0 \,\omega_0^2 \, \left(g(t+\tau)G(t)+g(t)G(t+\tau) \right)
\label{vcorrtau} \,   .
\nonumber \\
\end{eqnarray}

\end{widetext}


Note that the analytical expressions (\ref{xcorrtau}) and (\ref{vcorrtau})
are exact and  valid for all   absolute times $t$ and time lags $\tau$.
However, to evaluate  the  MSD (\ref{defmsd}) and VACF  (\ref{defvacf}) we must take the limit
$t\rightarrow\infty $. In this case,
these expressions could be simplified as follows.
Taking into account  the usual assumption  that the  time-dependent frictional
coefficient $\gamma(t)$
goes to zero when $ t \to \infty $ \cite{DV1} and  using the final value theorem \cite{Spig} one gets
\begin{eqnarray}
\lim_{ t \to \infty } \gamma( t)  = \lim_{ s \to 0 } s \widehat{
\gamma}(s) = 0 \,   .
 \label{condi1}
\end{eqnarray}
Noticing that the Laplace transform of the relaxation function $I(t)$ defined through Eq.
(\ref{IGs}) is
\begin{eqnarray}
\widehat{ I}(s) =  \frac{s^{-1}}{ s^2
+ s \, \widehat{ \gamma}(s) + \omega_0 ^2} \label{kernellades} \, ,
\end{eqnarray}
the application of the final value theorem and the use of condition (\ref{condi1}) yields \cite{DV1}
\begin{eqnarray}
I(\infty ) &=& 1/{\omega_0 ^2} \label{Iinfty} \, ,
\end{eqnarray}
and using (\ref{IGs}) and (\ref{Is}) gives
\begin{eqnarray}
G(\infty)&=& g(\infty)= 0   \label{ginfty} \, .
\end{eqnarray}

Applying  these conditions in order to take the limit $ t \to \infty $
in (\ref{xcorrtau}) and (\ref{vcorrtau}),  and using the definitions
(\ref{defmsd}) and (\ref{defvacf}) one finally obtain the simpler expressions
%
%
\begin{eqnarray}
\rho(\tau)=  \frac{2 k_B T}{m}
I(\tau)  \, ,
\label{msdfin}
\end{eqnarray}
and
\begin{eqnarray}
C_V(\tau)=  g(\tau)  \,   .
\label{cvfin}
\end{eqnarray}
%

Taking into account (\ref{Iinfty}) and (\ref{ginfty}), the equilibrium value of the MSD is given by
\begin{eqnarray}
\rho(\infty)=  \frac{2 k_B T}{m\omega_0 ^2}
 \, ,
\end{eqnarray}
while, as expected, the VACF decays to zero, i.e. $C_V(\infty)= 0$.

It is worth pointing out that in experimental realizations
the time lag is $\tau_\text{min}\leq\tau\leq\tau_\text{max}$, being $\tau_\text{min}$ the
acquisition time interval and $\tau_\text{max}$ the measurement time.
Moreover, if $N$ is the number of steps $n$ taken at intervals $\tau_\text{min}$,
only small values of $n$ $(n < N/10)$ are used.
Therefore, it is important to obtain valid expressions for all observational time scales
instead of getting only its behavior to large times.


To conclude this section, we will find the extension for a trapped particle
of the widely used Mason formula \cite{Ma95,Ma97}.
Taking the Laplace transform of (\ref{msdfin})
and using the definition (\ref{kernellades}) of the relaxation function $I(t)$, one gets
\begin{eqnarray}
s\widehat{ \gamma}(s) =  \frac{2 k_B T}{m} \frac{1}{ s \,\widehat{\rho}(s)}
-s^{2}- \omega_0 ^2 \, ,
\end{eqnarray}
which gives a direct relation between the mean-square displacement of the particle
and the memory kernel, from which the viscoelastic shear
moduli  of the medium can be obtained \cite{Wai}.


\section{Subdiffusive behavior}

Notice that the previous results are valid for any memory  kernel that
satisfy condition (\ref{condi1}).
On the other hand, it is well known that in the absence of active transport
the dynamics of the particle in a viscoelastic fluid or complex media is
subdiffusive and thus the stochastic process presents a long-time tail noise.
The most utilized model to reproduce a subdiffusive behavior is characterized
by a noise correlation function  exhibiting a power-law time decay \cite{Wa1,Lu2,VD}:
\begin{eqnarray}
 C(t) = C_{\lambda} \,\frac{ t^{-\lambda}}{\Gamma( 1 - \lambda)},
\label{correfracwang}
\end{eqnarray}
where $\Gamma(z)$  is the Gamma function \cite{Po}. The exponent $\lambda$ is taken as $0 < \lambda < 1$ and the
proportionality coefficient $C_\lambda$ is independent of time but
can depends on the exponent $\lambda$.

Using the fluctuation-dissipation relation (\ref{tfd}), the memory
kernel $\gamma(t)$ can be written as
\begin{eqnarray}
\gamma(t)=  \frac{\gamma_\lambda } {\Gamma(1-\lambda)}\,t^{- \lambda} \,
, \label{memdes2}
\end{eqnarray}
where $\gamma_\lambda = C_\lambda/k_B T $. Then, its Laplace transform reads
\begin{eqnarray}
 \widehat{ \gamma}(s) =  \gamma_{\lambda} \, s^{\lambda -1} \, .
 \label{kerfrac}
\end{eqnarray}

In this situation, the Laplace transform of the relaxation function $\widehat{ I}(s)$ reads
\begin{eqnarray}
\widehat{ I}(s) =    \frac{s^{-1}}{
s^2 + \gamma_{\lambda}\, s ^{\lambda} + \omega_0 ^2}
\label{kernellades2} \, .
\end{eqnarray}

The complete temporal behavior of the relaxation functions $I(t)$, $G(t)$ and $g(t)$ was previously obtained by us
in Ref. \cite{VD}.
Using those results  in (\ref{msdfin}) and (\ref{cvfin}) we have
\begin{eqnarray}
\rho(\tau) &=&  \frac{2 k_B T}{m} \, \sum_{k = 0}^{\infty}  \frac{(-1)^k}{k!} \, (\omega_0 \tau)^{2
k} \tau^{2}
\nonumber \\
&& \qquad \qquad \times \, E_{2-\lambda, 3 + \lambda k}^{(k)} (-\gamma_\lambda
\,\tau^{2- \lambda})
\label{msdsd}
\, ,  \\
C_V(\tau) &=& \sum_{k = 0}^{\infty}  \frac{(-1)^k}{k!} \, (\omega_0
\tau)^{2k}
\nonumber \\
&& \qquad \qquad
\times \, E_{2-\lambda,1 + \lambda k }^{(k)} (-\gamma_\lambda
\,\tau^{2- \lambda}) \, ,  \label{cvsd}
\end{eqnarray}
where $E_{\alpha, \beta} (y)$ is the generalized Mittag-Leffler
function \cite{Po}  defined by the series expansion
\begin{eqnarray}
E_{\alpha, \beta}(y) = \sum_{j = 0}^{\infty}  \frac{y^{j}}{\Gamma(
\alpha j + \beta)}    , \quad \alpha > 0, \quad \beta > 0 \, ,
\label{mittaglef}
\end{eqnarray}
and $E_{\alpha,\beta}^{(k)} (y)$ is the derivative of the Mittag-Leffler
function
\begin{eqnarray}
E_{\alpha,\beta}^{(k)} (y) =  \frac{ d^k}{dy^k}E_{\alpha,\beta}(y)
= \sum_{j = 0}^{\infty} \frac{( j + k )! \, y^{j}}{j ! \,\Gamma(
\alpha ( j + k ) + \beta)} \, . \label{demittaglef}
\end{eqnarray}



Using the series expansions (\ref{mittaglef}) and  (\ref{demittaglef}) one can realize
that the short times behavior  of the MSD  reads
\begin{eqnarray}
\rho(\tau) & \approx &  \frac{k_B T}{m} \,
\left\{\tau^{2}  - \frac{2\gamma_\lambda }{\Gamma(5-\lambda)} \,\tau^{4-\lambda}
- \frac{\omega_0^{2}}{12} \,\tau^{4}\right\} \, ,
\end{eqnarray}
where  the first term shows  that the particle undergoes ballistic motion when time is very small \cite{Wa2}.
The second term comes from the influence of the viscoelastic medium while the third term  corresponds
to the fact that the particle begins to ``see"  the trap.
The short times behavior  of the VACF can be obtained in a similar way. In this case we get
\begin{eqnarray}
C_V(\tau) & \approx & 1- \frac{\gamma_\lambda }{\Gamma(3-\lambda)} \,\tau^{2-\lambda}
- \frac{\omega_0^{2}}{2} \,\tau^{2}
\, .
\end{eqnarray}
%


On the other hand, for $\gamma_\lambda
\,\tau^{2- \lambda}\gg 1$ the MSD and VACF can be obtained
introducing the asymptotic behavior of the Mittag-Leffler function \cite{Po},
\begin{eqnarray}
E_{\alpha, \beta}(-y) \thicksim   \frac{1}{ y \, \Gamma( \beta-
\alpha)} \, ,  \quad    y > 0
\label{mitaglfpas}
\end{eqnarray}
into Eqs. (\ref{msdsd}) and (\ref{cvsd}). After some calculations  we have
\begin{eqnarray}
\rho(\tau) & \approx & \frac{2 k_B T}{m\omega_0^2}
\left\{1  -  \, E_{\lambda} ( -
\frac{\omega_0^2 }{ \gamma_\lambda}\, \tau ^{\lambda}) \right\} \, , \label{msdap}
\\
C_V(\tau) & \approx & - \frac{1} {\omega_0^2} \, \frac{d^2}
{d^2t}E_{\lambda} ( - \frac{\omega_0^2 }{ \gamma_\lambda}\, \tau ^{\lambda}) \, ,
\label{cvap}
\end{eqnarray}
where $E_{\lambda}(y)=E_{\lambda,1}(y)$ denotes the one parameter Mittag-Leffler function \cite{Po}.


It is worth pointing out that these expressions can be also obtained discarding
the inertial term $s^{2}$ in (\ref{kernellades2}).
In this case we get
\begin{eqnarray}
\widehat{ I}(s) =  \frac{s^{-1}}{
 \gamma_{\lambda}\, s ^{\lambda} + \omega_0 ^2} = \frac{1}{\omega_0 ^2} \left(\frac{1}{s}-\frac{s ^{\lambda-1}}{s ^{\lambda}+\omega_0 ^2/\gamma_{\lambda}}\right)
 \, ,
\end{eqnarray}
and  using that the Laplace transform  of the  Mittag-Leffler function \cite{Po}
\begin{eqnarray}
\int_{0}^{\infty} \, e^{-st} \,E_{\alpha}(-\gamma t^{\alpha}) \, dt = \frac{ s^{ \alpha -
1}}{s^{\alpha} + \gamma }  \,  ,
\label{derimittaglefdosparalaplacekcero}
\end{eqnarray}
one obtains expressions (\ref{msdap}) and  (\ref{cvap}).


Finally, if $ \tau ^{\lambda}\gg \gamma_\lambda/\omega_0^2$
the behavior of the MSD and VACF  can be obtained
using again the approximation (\ref{mitaglfpas}). In this case we get
\begin{eqnarray}
 \rho(\tau)&\approx& \frac{2 k_B T}{m\omega_0^2}\left\{1 - \frac{{\gamma_\lambda}} {\omega_0^2} \, \frac{1 }{\Gamma(1-\lambda) }
 \,\tau ^{-\lambda }\right\}\, , \label{msdas}\\
C_V(\tau) &\approx&  -\frac{{\gamma_\lambda}} {\omega_0 ^4}
 \, \frac{\lambda (\lambda+1)
 }{ \Gamma(1-\lambda) } \, \tau ^{-(\lambda + 2) }\, ,
 \label{cvas}
\end{eqnarray}
showing a pure power law decay.


\begin{figure}
\begin{center}
\includegraphics[scale=.6]{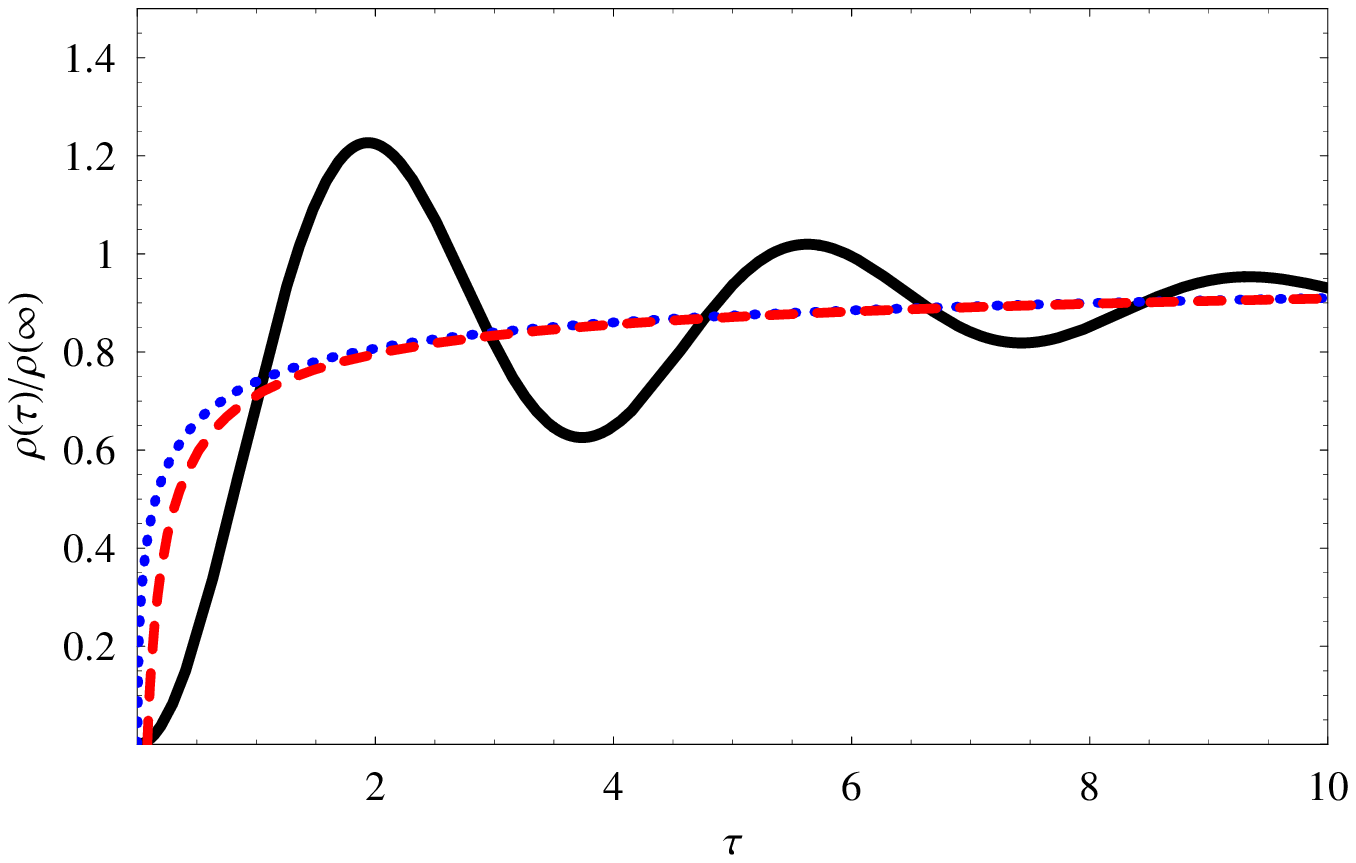}
\end{center}
\caption{ (color online) MSD vs. time lag   for
$\lambda = 1/2$, $\gamma_\lambda=1$ and $\omega_0=1.4$.
The solid line corresponds to the exact solution (\ref{msdsd}),
the dotted line to the approximate solution (\ref{msdap})
and the  dashed line to the asymptotic behavior (\ref{msdas}).} \label{fig1}
\end{figure}
\begin{figure}
\begin{center}
\includegraphics[scale=.6]{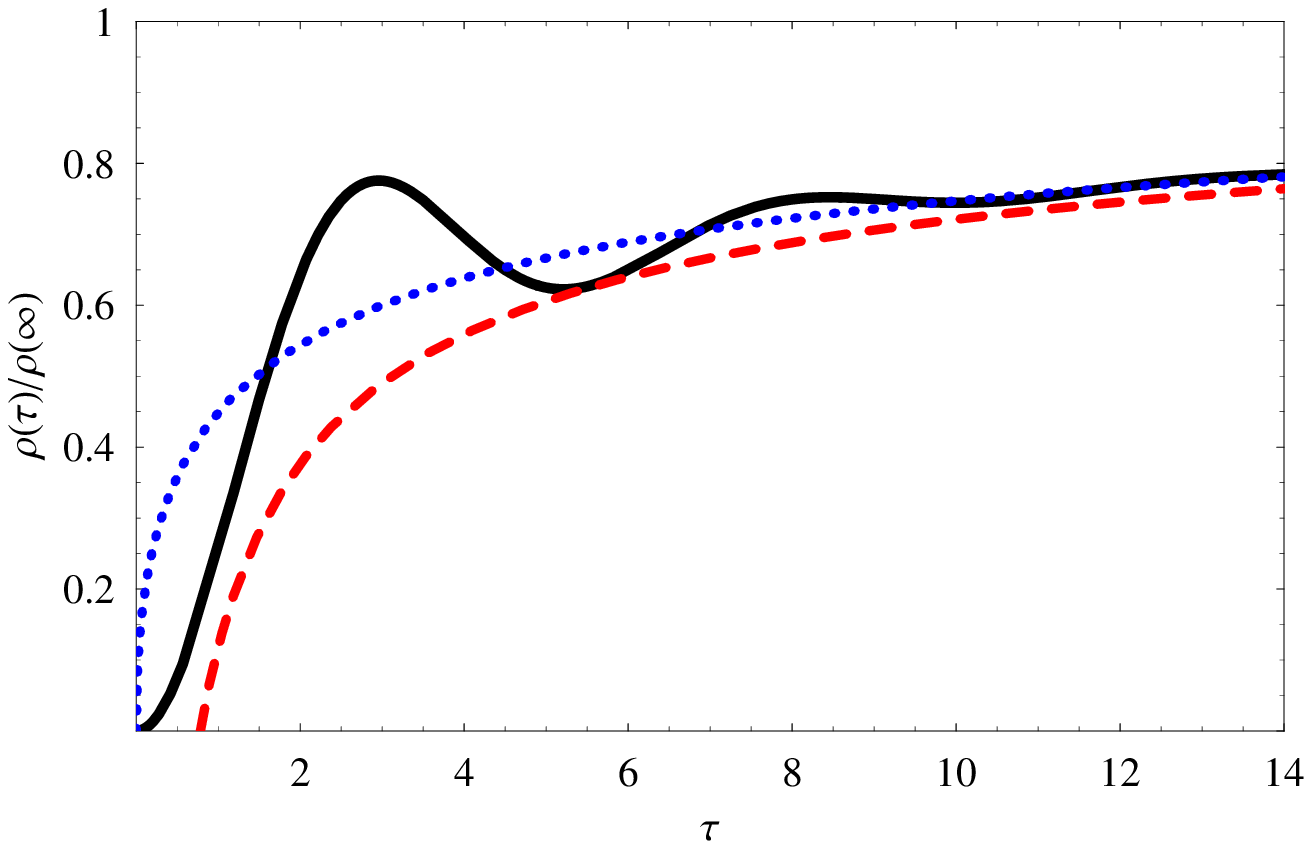}
\end{center}
\caption{ (color online) Idem Fig. \ref{fig1} for $\omega_0=0.8$. } \label{fig2}
\end{figure}

\begin{figure}
\begin{center}
\includegraphics[scale=.6]{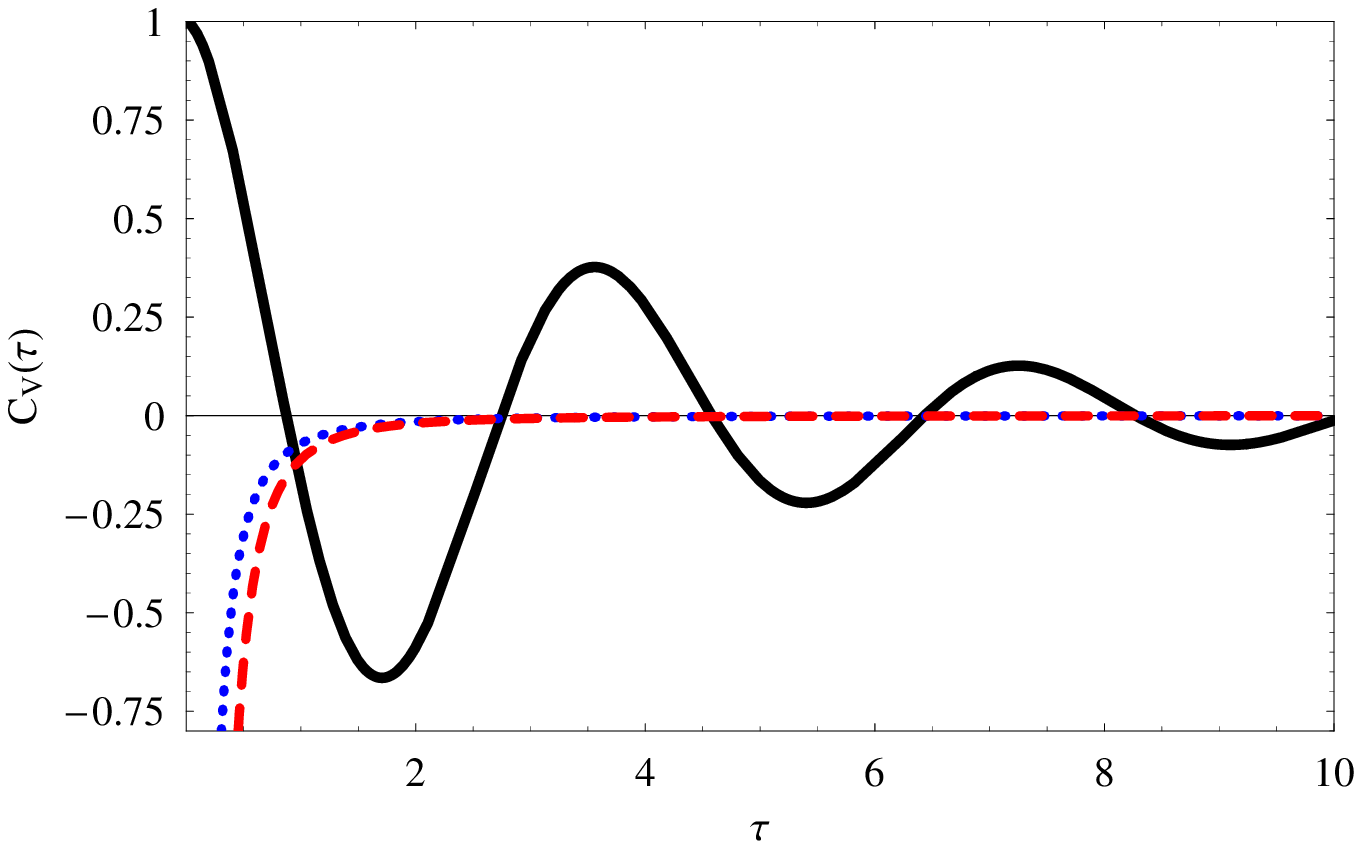}
\end{center}
\caption{ (color online)  $C_V$ vs. time lag   for
$\lambda = 1/2$, $\gamma_\lambda=1$ and $\omega_0=1.4$.
The solid line corresponds to the exact solution (\ref{cvsd}),
the dotted line to the approximate solution (\ref{cvap})
and the  dashed line to the asymptotic behavior (\ref{cvas}).} \label{fig3}
\end{figure}

\begin{figure}
\begin{center}
\includegraphics[scale=.6]{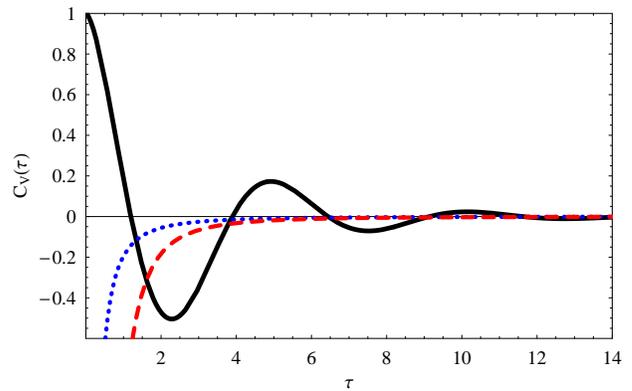}
\end{center}
\caption{(color online) Idem Fig. \ref{fig3} for $\omega_0=0.8$ .} \label{fig4}
\end{figure}


In Figs. \ref{fig1} and \ref{fig2} we have plotted  the MSD  vs. time lag,
using the exact solution (\ref{msdsd}) and the approximations (\ref{msdap}) and (\ref{msdas}).
Note that the exact solution exhibits a nonmonotonic
approach to $\rho(\infty)$, while the  approximations always present a monotonic behavior.
Furthermore, even in the overdamped case  the exact solution presents oscillations.
The same differences  can be observed in the behavior of $C_V(\tau)$,  as is  evidenced
in Figs. \ref{fig3} and  \ref{fig4}.
Interestingly,  Burov and Barkai \cite{Bark} have recently arrived to similar conclusions
examining the behavior of the position correlation $\langle X(t) X(0) \rangle$.

These behavior can be understood taking into account that the approximations
(\ref{msdap}) and (\ref{cvap}) only depend on the
one parameter Mittag-Leffler function
$E_{\lambda} ( -\omega_0^2 \, \tau ^{\lambda}/\gamma_\lambda) $.
On the other hand, it is known that the function
$E_{\lambda} (-t^{\lambda})$ is a completely monotone function and tends to zero
from above as t tends to infinity for $0< \lambda < 1$ \cite{Go}.
Then, the approximate solutions are  always
monotonic for every value of $\omega_0$  and  $0< \lambda < 1$.
However, the exact solutions (\ref{msdsd}) and (\ref{cvsd}) are expressed as
infinite sums of $E_{2-\lambda, \beta}^{(k)} (-\gamma_\lambda
\,\tau^{2- \lambda})$ functions.
In this case, the solutions can exhibit a nonmonotonic behavior as is displayed in the previous figures.


\section{Summary}

In this work we have obtained  the mean square displacement and
the velocity autocorrelation function for a trapped particle
and immersed in a complex or viscoelastic media.
For this purpose, and starting from a suitable generalized Langevin equation,
we have been able to derive analytic expressions for the two-times dynamics of
the processes, valid for all absolute times and times lags.
We have showed that the MSD and VACF can be expressed as a simple expressions
when the memory kernel goes to zero for large times.
In particular, we have examined the subdiffusive case,
the which one is paradigmatic in the study of passive transport in viscoelastic media.
In this case, exact expressions and valid for all time lags
have been obtained in terms of Mittag-Leffler functions and its derivatives.
The limit of short time lags are given in terms of the involved parameters.
Finally, we have showed that the overdamped approximation, which means that the effects
of inertia are neglected, can not reproduce the nonmonotonic dynamics present in  the exact solutions.
This result must be taken into account in the analysis of the short and intermediate
times dynamics where the MSD and VACF exhibit   a relaxation plus an oscillatory behavior.

In summary, we have presented a method to account for the
effects of the trapping potential in the anomalous behavior of the
mean square displacement
and the normalized velocity autocorrelation function
of a particle embedded in a complex or viscoelastic environment.


\begin{acknowledgments}

This work was performed under Grant N$^\circ$  PICT
31980/05 from Agencia Nacional de Promoci\'{o}n Cient\'{i}fica y
Tecnol\'{o}gica, and Grant N$^\circ$ X099 from Universidad de Buenos Aires,
Argentina.

\end{acknowledgments}


\appendix*

\section{\label{ApA} }

To calculate the last
term of Eqs. (\ref{Xss}) and (\ref{Vss})  we make use
a relation given in Ref.\cite{Pot}. Given any
stationary correlation function of the form
\begin{eqnarray}
\langle \Psi(t) \Psi(t') \rangle = A f(|t - t'|)
 \, ,\label{corregene}
\end{eqnarray}
the corresponding  double Laplace transform writes
\begin{eqnarray}
\langle \widehat{ \Psi}(s) \widehat{ \Psi}(s') \rangle = A \frac{
 \widehat{f}(s) +  \widehat{f}(s')}{ s + s'}
 \, .\label{reladfdlg}
\end{eqnarray}
Then, the Laplace domain version of  the fluctuation-dissipation relation
(\ref{tfd}) reads \cite{Pot}
\begin{eqnarray}
\langle  \widehat{F}(s)\widehat{F}(s') \rangle = k_B T m\,
\frac{\widehat{\gamma}(s)+\widehat{\gamma}(s')}{s+s'}
\label{fdss} \,   .
\end{eqnarray}

After some algebra, and using the relations between the kernels
$I(t)$ $G(t)$ and $g(t)$  one can  find that
%
%
\begin{widetext}
%
\begin{eqnarray}
\widehat{G}(s)\widehat{G}(s')
\langle \widehat{F}(s)\widehat{F}(s')\rangle
&= &k_B T m\,\left(
\frac{\widehat{I}(s)}{s'}
+ \frac{\widehat{I}(s')}{s}-\frac{\widehat{I}(s)+\widehat{I}(s')}{s+s'}\right)
-k_B T  m \,\left(\widehat{G}(s)\widehat{G}(s')+\omega_0^2\, \widehat{I}(s)\widehat{I}(s')\right)
\,     ,    \label{GGFF} \\
\widehat{g}(s)\widehat{g}(s')\langle
  \widehat{F}(s)\widehat{F}(s') \rangle &= &
k_B T m\left(\frac{\widehat{g}(s)+\widehat{g}(s')}{s+s'}
 - \widehat{g}(s)\,\widehat{g}(s')- \omega_0^2\, \widehat{G}(s)\widehat{G}(s')\right)
  \,    .\label{ggFF}
\end{eqnarray}
%
\end{widetext}



\end{document}